\documentclass[a4paper]{jpconf}
\usepackage{iopams}
\begin{document}

\title{Relativistic state reduction model}
\author{Daniel Bedingham} 
\address{Blackett Laboratory, Imperial College, London, SW7 2BZ}
\ead{d.bedingham@imperial.ac.uk}

\begin{abstract}
In order to address the measurement problem of quantum theory we make the assumption that 
quantum state reduction should be regarded as a genuine physical process deserving of a 
dynamical description. Generalizing the nonrelativistic spontaneous localization models of 
Ghirardi, Rimini, Weber, and Pearle, a relativistic state reduction mechanism is proposed. 
The mechanism involves nonlinear stochastic modifications to the standard description of 
unitary state evolution and the introduction of a mediating field to facilitate smearing 
of quantum field interactions.
\end{abstract}

\section{Introduction}

Here we shall address the question of how to understand quantum states and their evolution. The conventional view is that
the quantum state is a complete description of a system. The state at some time can be determined from the state at some earlier
time by using the Schr\"odinger equation. {\it That is}, unless there is a measurement, at which point a quantum state reduction occurs: 
the state changes in a discontinuous and stochastic way consistent with the observed measurement outcome. 
There is no underlying theory to decide precisely when a measurement occurs in this picture and the fact that such a fuzzy concept 
plays such a key role is the quantum measurement problem.

In an attempt to address this issue we shall make two assumptions: (i) the quantum state is a real physical structure (in principle it
has some definite value at any given stage despite our possible inability to determine it accurately), and (ii) both the Schr\"odinger 
equation and quantum state reduction describe real physical processes which continually affect the quantum state (each to 
a greater or lesser extent at any given stage). 

The task is now to find a set of dynamical equations, more general than the Schr\"odinger equation, which 
can be approximated by either the Schr\"odinger equation or quantum state reduction in situations where each of those descriptions is 
appropriate. In this way we avoid having to place the concept of measurement at the centre of our theory.

In particular we shall address the question of how to do all this whilst satisfying the constraints of relativity. After a brief 
review of some existing nonrelativistic models we shall outline our proposal for a relativistic state reduction mechanism.
(For a general review of dynamical state reduction models see ref.~\cite{Bass}.)

\section{Nonrelativistic state reduction models}
\subsection{Spontaneous localization of particles}
We begin by reviewing the spontaneous localization model of Ghirardi, Rimini, and Weber \cite{ghir3}. The model concerns the quantum 
mechanics of a system of distinguishable particles. 

Consider this system described at time $t$ by the state $|\psi(t)\rangle$. For each particle we consider a sequence of Poisson distributed
random points in time (with frequency of order $10^{-17}$s$^{-1}$ making these events extremely rare). 
At each of these random times the state changes discontinuously under the action of a localization operator (e.g., for particle $i$)
\begin{equation}
|\psi(t)\rangle \rightarrow |\psi(t+)\rangle = L_i({\bf z})|\psi(t)\rangle.
\end{equation}
At all other times the state evolution satisfies the Schr\"odinger equation. The localization operator is defined by
\begin{equation}
L_i({\bf z})= \frac{1}{(\pi r^2)^{3/4}} e^{-\frac{({\bf q}_i- {\bf z})^2}{2 r ^2}},
\label{collapse}
\end{equation}
where ${\bf q}_i$ is the (3D) position operator for particle $i$ and ${\bf z}$ is a real random variable representing a random point in position 
space. The operator $L_i({\bf z})$ has the effect of focusing the quantum amplitude (in position space) for particle $i$ about the point 
${\bf z}$. The parameter $r$ is of order $10^{-7}$m and determines the length scale at which localizations occur. 

The localization centre ${\bf z}$ for particle $i$ is chosen from a probability distribution defined by
\begin{equation}
\mathbb{P}_i({\bf z}\in D) 
= \int_D d{\bf z} \; \frac{\langle\psi(t)|L^2_i({\bf z})|\psi(t)\rangle}{\langle\psi(t)|\psi(t)\rangle},
\label{probrule}
\end{equation}
for some spatial region $D$. This probability rule ensures that the statistical connection between the quantum state, pre and post
localization is maintained (i.e.~the Born rule is satisfied).

Now consider a lump of matter composed of vast numbers of particles. Although the chance of a localization occurring in a small amount 
of time for any one particle is very small, for a sufficiently large number of particles, the chance for at least one particle to be 
localized can be large. Entanglements between particle positions then entail that bulk superpositions are suppressed. This is the
mechanism by which state reductions occur and it explains why they can be neglected at the scale of individual particles. We must be 
careful to understand here that a measurement is caused by the occurrence of a localization event, not the other way around.

\subsection{Spontaneous localization with fields}
\label{basis}
The major drawback for the spontaneous localization model outlined above is that all particles must be distinguishable. If we allow
particles to be indistinguishable then the state must satisfy certain symmetries under particle interchange. If we continue to consider 
localization events occurring for individual particles then they will break these symmetries.

The resolution of this dilemma, as suggested by Ghirardi, Pearle, and Rimini \cite{ghir2}, is to treat the system of particles in
the manner of a quantum field.  By this we mean that we consider annihilation and creation operators $a({\bf x})$ and 
$a^{\dagger}({\bf x})$ which respectively describe the annihilation and creation of particles at point ${\bf x}$ 
(the field operator at ${\bf x}$ is $a({\bf x})+a^{\dagger}({\bf x})$ and 
$[a({\bf x}),a^{\dagger}({\bf y})]=\delta^3({\bf x}-{\bf y})$). This enables us to define a smeared number 
density operator by
\begin{equation}
N({\bf x})=\int d{\bf y} \; g({\bf x}-{\bf y}) a^{\dagger}({\bf y})a({\bf y}),
\end{equation}
where $g({\bf x})$ is a spherically symmetric, positive real function peaked around ${\bf x} = {\bf 0}$.

Now consider points Poisson distributed in {\it spacetime volume} with a certain fixed frequency density independent of the system. 
The state evolution satisfies the Schr\"odinger equation except for those times when the timeslice (in spacetime) crosses one of the 
random points. At these times the state is acted upon by the localization operator
\begin{equation}
L(Z_x)= \frac{1}{(\pi r^2)^{3/4}} e^{-\frac{\left(N({\bf x})- Z_x\right)^2}{2 r^2}},
\label{collapZ}
\end{equation}
where $x=({\bf x},t)$ is the random spacetime point and $Z_x$ is a real positive random variable representing a preferred particle number 
density at $x$. The effect of the operator $L(Z_x)$ is to resolve different number density states at ${\bf x}$ and focus the quantum 
amplitude about $N({\bf x})=Z_x$. The probability rule for $Z_x$ is
\begin{equation}
\mathbb{P}(Z_x\in D_x) 
= \int_{D_x} d Z_x \; \frac{\langle\psi(t)|L^2(Z_x)|\psi(t)\rangle}{\langle\psi(t)|\psi(t)\rangle},
\end{equation}
where $D_x$ is some range of possible $Z_x$ values.
The localizations are now labelled by spacetime points $x$ rather than particle index $i$ and time.

This picture is quite different from before. Since the random distribution of points in spacetime has nothing to do with the
quantum system itself, the points must be quite dense (to account for the possibility of state reductions occurring at any time). 
But in order to ensure that individual particles are not continuously being localized, the effect of the localization operator
must be tuned such that it is only effective for states involving a large spread in $N({\bf x})$ values ($r$ must be sufficiently large). 
Such states correspond to bulk superpositions.

The smearing function $g({\bf x})$ is necessary since without it the action of the
localization operator would result in spatial discontinuities in the number density of particles. Such states have infinite energy. 
The length scale associated with $g$ is in fact the length scale associated with the localizing behaviour of this model.

Note that in the model presented in ref.~\cite{ghir2}, a limit is taken in which the localization operator acts at every single point
(giving a continuous localization process). Here for simplicity we consider only discrete localization events.

The developments outlined here pave the way toward a relativistic formulation. Although the 
state is still described only at a sequence of times (in some frame of reference), the localization events (or hits) 
have an invariant distribution in spacetime (a Poisson distribution in spacetime volume is invariant since spacetime volumes are invariant). 
The challenge of formulating a fully relativistic model is to understand how the state moves through the invariant minefield of hits 
in a way which does not refer to any preferred timeslice.

\section{Relativistic state reduction}
\subsection{Tomonaga picture}
In the Tomonaga picture \cite{Tomo} a quantum state is assigned, not just to a sequence of timeslices in some preferred frame, but to every
spacelike hypersurface $\sigma$. In this way we do not discriminate between different frames of reference, neither locally nor globally. Given an interaction
Hamiltonian density $H_{\rm int}(x)$, the dynamics of the state is given by the Tomonaga equation
\begin{equation} 
i \frac{\delta}{\delta\sigma(x)}|\Psi(\sigma)\rangle=H_{\rm int}(x)|\Psi(\sigma)\rangle,
\end{equation}
where
\begin{equation}
\frac{\delta|\Psi(\sigma)\rangle}{\delta\sigma(x)}=
\lim_{\sigma'\rightarrow\sigma}\frac{|\Psi(\sigma')\rangle-|\Psi(\sigma)\rangle}{\Delta\omega},
\end{equation}
and where $\Delta\omega$ is the spacetime volume enclosed by $\sigma$ and $\sigma'$ about point $x$ (with no points in $\sigma'$ 
to the past of $\sigma$). In order that the Tomonaga equation has covariant form the Hamiltonian density $H_{\rm int}(x)$ must be a Lorentz 
scalar. Furthermore, it must be the case that $[H_{\rm int}(x),H_{\rm int}(y)]=0$ for spacelike separated $x$ and $y$. This 
rule means that there is no preferred ordering of spacelike separated points and ensures a consistent and frame-independent 
assignment of states to hypersurfaces.

\subsection{Relativistic hits}

The solution to the Tomonaga equation is given by
\begin{equation}
|\Psi(\sigma_1)\rangle \rightarrow |\Psi(\sigma_2)\rangle = 
T e^{-i\int_{\sigma_1}^{\sigma_2}d^4 x \; H_{\rm int}(x)} |\Psi(\sigma_1)\rangle,
\end{equation}
with no points in $\sigma_2$ to the past of $\sigma_1$ and where $T$ describes path ordering of timelike separated events. This 
solution represents the unitary evolution of the quantum state in spacetime.
Our goal is to supplement this background with random localization events or hits. As in section \ref{basis} we consider an invariant Poisson
distribution of points in spacetime with fixed frequency density independent of the system. 

Consider advancing the hypersurface $\sigma$ in a timelike direction. We assume that when $\sigma$ crosses one of the random spacetime points 
(at $x$ say), in addition to the unitary evolution, the state changes by
\begin{equation}
|\Psi(\sigma)\rangle \rightarrow |\Psi(\sigma_+)\rangle = L(Z_x)|\Psi(\sigma)\rangle.
\label{relhit}
\end{equation}
We suppose that $L(Z_x)$ is given by 
\begin{equation}
L(Z_x)= \frac{1}{(\pi r^2)^{3/4}} e^{-\frac{\left(N(x)- Z_x\right)^2}{2 r^2}},
\end{equation}
and base the following discussion around determining a possible
form for relativistic version of the operator $N(x)$.

First of all $N(x)$ must be a Lorentz scalar in order that the (stochastic) equations of motion have covariant form (just as for
$H_{\rm int}$). Next we must have
\begin{equation}
[N(x),N(y)]=0 \quad {\rm and} \quad [N(x),H_{\rm int}(y)]=0,
\label{Nconst}
\end{equation}
for any spacelike separated $x$ and $y$ (meaning no preferred ordering of spacelike separated points).

The first obvious model to consider is a scalar field $\phi$ where we take $N=\phi$ (with $H_{\rm int}$ possibly 
equal to $\phi^4$ or some interaction between $\phi$ and other fields) \cite{pear3,pearGhir}. This choice satisfies (\ref{Nconst}), 
but if we attempt to calculate the expected change in energy as a hit point is traversed by $\sigma$ we find an infinite result. The reason is 
because the pointlike hits lead to spatial discontinuities in the $\phi$ state of the field. It is for this same reason that a smearing 
function $g$ was employed in section \ref{basis}. We must try to find a invariant {\it smeared} operator $N$ for which (\ref{Nconst}) 
still holds.

\subsection{Mediating field}
\label{medsec}
Here we suggest a solution to this problem making use of a mediating field \cite{ME}.
The mediating field we use in our state reduction mechanism has some unconventional properties designed to enable us to 
fulfill the constraints outlined above. We define annihilation and creation operators $a(x)$ and $a^{\dagger}(x)$ with the 
property
\begin{equation}
[a(x),a^{\dagger}(y)] = \delta^4(x-y).
\end{equation}
Field states are defined by repeated application of the particle creation operator on the vacuum state $|0\rangle$ 
(which the annihilation operator annihilates). For example, a first excited state is given by
\begin{equation}
|h\rangle =  \int d^4 x \; h(x) a^{\dagger}(x)|0\rangle,
\end{equation}
for some square integrable function $h$. Note that whereas the modes of a conventional quantum field describe a 
field configuration on a timeslice (or spacelike hypersurface), the degrees of freedom of the mediating field describe 
a field configuration over the whole of spacetime.

The point of constructing a field like this is that it enables us to specify a smeared number density operator satisfying 
(\ref{Nconst}). We define the number density operator by $n(x)=a^{\dagger}(x)a(x)$. We then define our smeared
number density as
\begin{equation}
N(x) = \int d^4 y \; f(x,y) n(y).
\end{equation}
We also define a smeared field operator by
\begin{equation}
A(x) = \int d^4 y \; g(x,y) \left[ a(y) + a^{\dagger}(y) \right].
\end{equation}
We do not specify $f$ and $g$ as yet but assume for now that they are invariantly specified and sufficiently well 
behaved functions.

On its own the mediating field is static; it has no intrinsic dynamics. Its dynamics must result from interaction with other 
quantum fields. For this we propose 
an interaction Hamiltonian density of the form $H_{\rm int}(x)=J(x)A(x)$ where $J(x)$ is some (Lorentz scalar) matter density 
operator for a conventional quantum field (e.g., $J=\phi^2$). This interaction has the effect of exciting the mediating 
field in proportion to $J(x)$. To clarify the picture we will assume that the mediating field is initially in its vacuum state
and that its only excitations are those caused by this interaction.

For all $x$ and $x'$ the smeared operators have the commutation properties $[N(x),N(x')]=0$, $[A(x),A(x')]=0$, and
\begin{equation}
[N(x),A(x')]=\int d^4y \; f(x,y)g(x',y)\left[a^{\dagger}(y)-a(y)\right].
\end{equation}
This means that (\ref{Nconst}) is satisfied if (i) $f(x,y)$ is only nonzero for $y$ the causal past of $x$, and 
(ii) $g(x,y)$ is only nonzero for $y$ the causal future of $x$. In this case $[N(x),A(x')]=0$ for spacelike separated $x$
and $x'$ (and consequently  $[N(x),H_{\rm int}(x')]=0$ for spacelike separated $x$ and $x'$).

Before summarizing our relativistic state reduction mechanism we remark on the functions $f$ and $g$. These
functions may depend on any Lorentz invariant local properties of the model (e.g, local operator expectations invariantly 
specified by the state assigned to the past light cone hypersurface; realized values of random variables $Z_x$, see \cite{ME}). 
This enables us to find locally preferred frames invariantly determined by the state (e.g., its local rest frame), with 
respect to which the localizations occur. Without this feature we would not be able to talk about localization.

\subsection{State reduction mechanism outline}
With the mediating field initially in the ground state consider a quantum field in a superposition of two different 
matter density states ($J$ states). This might correspond to a superposition of two displaced lumps of matter. The 
subsequent interaction between $J(x)$ and $A(x)$ (described by the Tomonaga equation) leads to an excitation of the mediating 
field. The result is an entangled superposition in which each matter density state is associated with a mediating field state 
representing a smeared image of the path of matter through spacetime. This can be thought of as similar to the effect of a 
particle passing through a cloud chamber where a record of the track is formed and left behind.

Next suppose that a random hit occurs at some spacetime point where the mediating field has been excited. If the mediating
field is in a superposition of different $N(x)$ states at this point (each corresponding to different parts of the entangled superposition)
then the localization operator will choose between the different possibilities by suppressing the quantum amplitude of states
with $N(x)$ different from $Z_x$. Consequently, one of the matter density states in the superposition will be suppressed.

The absence of any point interactions in this mechanism ensures that the result is finite \cite{ME}.

\subsection{Probabilities}
Suppose that a sprinkling of Poisson distributed points in spacetime is given and suppose that we are interested in
the probabilities for $\{Z_x|\sigma_1<x<\sigma_2\}$ (i.e.~random variables $Z_x$ at all random points enclosed by the surfaces
$\sigma_1$ and $\sigma_2$). The probability that each $Z_x$ is in the range $D_x$ is given by
\begin{equation}
\mathbb{P}\left(\{Z_x\in D_x| \sigma_1< x <\sigma_2 \}\right)= \left(\prod_{\sigma_1< x <\sigma_2} \int_{D_x} dZ_x\right)\; 
\frac{\langle \Psi(\sigma_2)|\Psi(\sigma_2)\rangle}{\langle \Psi(\sigma_1)|\Psi(\sigma_1) \rangle},
\label{gprob}
\end{equation}
where it is understood that $|\Psi(\sigma_2)\rangle$ depends on $\{Z_x|\sigma_1<x<\sigma_2\}$ via equation (\ref{relhit}).

\subsection{State reduction timescale}
Consider an initial equal superposition of two displaced $J(x)$-states, each describing some localized lump of matter. 
We assume that each $J(x)$-state has some finite spatial extent (outside of which we can take $J(x)=0$).
Working in some fixed frame, let the spatial volume of points belonging to one but not both of the spatial extents of the 
two $J(x)$-states be $V_{\triangle}$. The state reduction timescale is then given (in this frame) by \cite{ME}
\begin{equation}
\tau \sim r^{2}\mu^{-1}V_{\triangle}^{-1}J^{-4},
\end{equation}
where $\mu$ is the frequency density of hit points in spacetime and $J$ is the typical value of $J(x)$ in $V_{\triangle}$.
This is a result which can in principle be experimentally tested \cite{MIRR}.

\subsection{Nonlocality}
Any proposed completion of quantum theory must violate the concept of Bell locality. Here we briefly remark on the way in
which this model is nonlocal.

Consider a state which describes two spacelike separated subsystems of an entangled global system (such as in an EPR-type 
experiment). Suppose that each of these subsystems undergoes a state reduction (such as that involved in a spin measurement). 
The probability rule (\ref{gprob}) ensures not only that the outcomes of the state reduction for each subsystem occur individually with 
the correct quantum probabilities but also that the joint probabilities for outcomes in the two subsystems satisfy quantum 
predictions. This implies that $Z_x$ values are correlated over spacelike separation in the physical probability measure. 
We might even think of $Z$ globally as a nonlocal hidden variable.

\section{Summary}
We have outlined a relativistic mechanism describing quantum state reduction in which, by construction, 
outcomes have the probabilities predicted by standard quantum theory. This has been achieved without the need to talk about 
measurements. Indeed within this framework, measurements are just special examples of localization phenomena.

The model offers the possibility of consistently and objectively explaining the behaviour of micro and macro objects. The 
mechanism can be applied to any quantum field for which we can form a Lorentz scalar operator $J(x)$ (both bosons and fermions), 
and is compatible with gauge field interactions.

The key to this model is the use of the mediating field which enables interactions to be smeared without reference to a preferred
foliation of spacetime.

\end{document}